\documentclass[12pt,preprint]{aastex}

\usepackage{natbib}
\usepackage{graphicx}
\usepackage{bbm}
\newcommand{\beq}{\begin{equation}}
\newcommand{\eeq}{\end{equation}}
\newcommand{\lsi}{\,\raisebox{-0.13cm}{$\stackrel{\textstyle<}
{\textstyle\sim}$}\,}
\newcommand{\gsi}{\,\raisebox{-0.13cm}{$\stackrel{\textstyle>}
{\textstyle\sim}$}\,}
\newcommand{\be}{\begin{equation}}
\newcommand{\ee}{\end{equation}}

\begin{document}

\title{Giant AGN Flares and Cosmic Ray Bursts}

\author{Glennys R. Farrar and Andrei Gruzinov}

\affil{Center for Cosmology and Particle Physics \&
Department of Physics\\ New York University, NY, NY 10003,USA\\
}

\keywords{cosmic rays, AGN, x-rays, GRB}

\begin{abstract}
We predict a new class of very intense, short-duration AGN flares capable of accelerating the highest energy cosmic rays, resulting from the tidal disruption of a star or from a disk instability.  The rate and power of these flares readily explains the observed flux and density statistics of UHECRs.  The photon bursts produced by the predicted AGN flares are discussed; they may soon be detectable.   Observations are shown to exclude that continuous jets of powerful Active Galactic Nuclei are the sole source of ultrahigh energy cosmic rays; the stringent requirements for Gamma Ray Bursts to be the source are delineated.
\end{abstract}

\section{Introduction}
A major challenge in theoretical High Energy Astrophysics has been to explain the mechanism by which ultrahigh energy cosmic rays are accelerated.  The difficulty stems from the almost contradictory requirements of the minimum magnetic field needed to confine and accelerate the cosmic rays, and the maximum magnetic field consistent with avoiding too much synchrotron radiation and photo-pion energy loss.  It was argued by \citet{waxman95} that Gamma Ray Bursts could overcome this near-contradiction -- if the relativistic gamma factors are sufficiently high -- but as we show below, UHECR acceleration by GRBs alone is difficult to reconcile with observational data.  Sufficiently powerful AGN jets and giant lobes of radio galaxies could also satisfy the requirements, but too few of requisite luminosity are observed within the GZK distance $\lsi 100$ Mpc.  

We argue here that due to tidal disruption of passing stars or episodic instabilities in the accretion disk, even weak AGN naturally have major flares with the required frequency and luminosity to account for the flux of UHE cosmic rays.   First, we review the basic requirements for UHECR acceleration.  Next, we determine the constraints from UHECR observations;  these pose a serious challenge to GRBs, and exclude steady state AGN jets, as the sole accelerators of UHECRs.  Then we develop the giant AGN flare scenario and calculate the frequency and luminosity expected for such flares; the implied UHECR flux is consistent with observation.  We discuss the properties of the photon bursts and show that GLAST and future optical surveys can test our predictions. 
\section{Parameters of UHECR accelerators}

\label{params}
The most obvious candidates for UHECR accelerators (e.g., GRB, powerful AGN) all have relativistic jets.  Without specializing to any particular system, we consider an outflow (jet) from some central engine, with mean Lorentz factor $\Gamma$ and with variations of the Lorentz factor $\sim \Gamma$.  Due to variability in the jet velocity, one gets a turbulent flow with (at least) mildly relativistic shocks -- as seen in the jet frame.  A similar analysis applies to the termination shock created when such an outflow enters the ISM.  We consider here only the acceleration of protons and we do not consider non-relativistic acceleration scenarios.

An estimate of the parameters required to accelerate UHECRs can be obtained as follows\footnote{See \citet{waxman95} for a clear discussion.}.  Let the characteristic size of the acceleration region in the jet frame be $\sim R$, with a characteristic magnetic field strength $\sim B$ (also measured in the jet frame). 
In the jet's frame, we require that the shock/turbulence accelerates a cosmic ray to energy $\sim E_{20} \,10^{20}\, \Gamma ^{-1}$eV.   To do so, the turbulent cloud must confine the cosmic ray, so
\begin{equation}\label{conf}
RB\gsi 3\times 10^{17} \,\Gamma ^{-1} 
\, E_{20}.  
\end{equation}
(Here and below we use cgs units unless specified otherwise.)   Furthermore, synchrotron losses during a given acceleration cycle must not exceed the energy gained in the cycle. The minimum energy loss in a typical acceleration cycle is $\sim  \pi R_L/c$ times the synchroton power emitted by the CR in field $B$, resulting in the condition
\begin{equation}\label{sync}
B\lsi \, \Gamma ^2  
\, E_{20} ^{-2}.
\end{equation}

UHECR acceleration in turbulent shocks is accompanied by a flux of photons, which can be an important signature of the process.  The (isotropic equivalent) total power in the required magnetic field (Poynting luminosity) is of order
\begin{equation}\label{lumi}
L \sim {1\over 6}c\Gamma ^4B^2R^2\gsi 10^{45} \Gamma ^2\, E_{20}^2 \, {\rm erg/s}.
\end{equation}
If the energy in the magnetic field, protons and electrons is in equipartition, and the energy in electrons is emitted through synchrotron cooling in the time it takes the shock to pass through the magnetic cloud, a comparable luminosity is also emitted by electrons with jet-frame Lorentz factors $10^3\lsi \gamma _e\lsi  10^8 \, B ^{-\frac{1}{2}}$;  the lower bound is from assuming equipartiton with the mildly relativistic ions, and the upper bound is from synchrotron cooling of the electrons.  The spectra of protons and electrons (in equipartition) are expected to be roughly flat -- equal energy per logarithmic energy interval, leading to an approximate photon spectrum\footnote{At best, equipartition is only roughly right but not accurate in detail, as evidenced by the GRB spectrum being far from flat.}
\begin{equation}
\label{spec}
\nu L_{\nu } \gsi 10^{44} \Gamma ^2 E_{20}^2, ~~~0.01~\Gamma B~{\rm eV}\lsi h\nu \lsi 10 ~ \Gamma ~{\rm MeV},
\end{equation}
beamed into a cone of opening angle $\gtrsim \Gamma^{-1}$.

Photons with energy above the threshold $\epsilon_\gamma \approx  100 \, {\rm MeV }/ (10^{11} \Gamma^{-1} E_{20}) \approx 10^{-15} E_{20}^{-1} \Gamma$ erg, have a cross section $\sigma_\pi \sim 10^{-28} \,{\rm cm}^2$ for a photo-pion production interaction, each of which typically decreases the proton energy by $\sim$10\%.  Therefore, the condition that photo-pion losses do not limit further acceleration is $n_\gamma \sigma_\pi R \lsi 10$, where $n_\gamma$ is the density of photons with energy above $\epsilon_\gamma$ and depends on the particular environment, cooling rate, etc.  We obtain an estimate of the maximum $n_\gamma$ by assuming the total jet-frame Poynting luminosity is carried by photons, distributed evenly over 10 logarithmic intervals.  In that case $n_\gamma \approx 10^{-3} \, B^2/\epsilon_\gamma$, which with (\ref{lumi}) gives a sufficient condition for avoiding excessive photo-pion energy losses:
\begin{equation}\label{phot}
RB^2 \lsi 10^{17} E_{20}^{-1} \Gamma.
\end{equation}
Although the precise coefficients in (\ref{conf},\ref{sync}) depend on the details of the turbulence and shocks, these constraints are quite general and should be robust at the level of a factor $\sim$few.  Conditions (\ref{spec}) and (\ref{phot}) rely in addition on the assumption of approximate equipartition of energy between electrons and protons and thus are less general.  

The Poynting luminosity required for UHECR acceleration in a relativistic jet (\ref{lumi}) need not translate to a lower bound on the luminosity emitted by the jet in photons since equipartition may not be valid.  Nonetheless, the total power -- (\ref{lumi}) times the beaming factor of the jet $\sim \Gamma^{-2}$ -- does provide a lower limit on the luminosity of the accretion disk which powers the jet.  This is not a theorem; indeed, if the disk provides a strong large scale magnetic field threading the black hole, the Blanford-Znayek mechanism \citep{bz77} can extract energy from the BH spin rather than from the accretion flow.   However such a low-entropy scheme would not be realized in any astrophysically realistic environment.  This can be seen explicitly using the classical Shakura-Sunyaev solution for a disk accreting at approximately the Eddington rate, to obtain the maximum magnetic field strength which can be contained by the pressure of the accretion disk.  Assuming the B-field at the BH attains this field strength with optimal direction, gives an upper limit on the maximum BZ luminosity, $L_{\rm BZ}$, which is $\approx$ the Shakura-Sunyaev accretion luminosity.  Therefore, $10^{45} E_{20}^{2}$ is a robust lower limit for the bolometric luminosity produced by the accretion disk powering a jet responsible for UHECR acceleration; beamed photon emission from the jet itself is in general also expected.

\section{Observational Constraints}

\label{obs}
A successful model for UHECR acceleration must be consistent with the following observational constraints:   \\
{ \it 1) Energy injection rate:} Estimates of the UHECR energy injection rate per logarithmic interval range from $0.7 -  20  \times 10^{44}  \, {\rm erg \, Mpc^{-3} \, yr^{-1}}$; the former is due to \citet{waxman95} and the latter from the fit of \citet{berez07}.  Thus define
\begin{equation} 
\label{Einj}
\Gamma_{\rm ElnE} = 10^{44.6 \pm 0.7} \Gamma_{44.6}\, {\rm erg \, Mpc^{-3} \, yr^{-1}}.
\end{equation}
{\it 2) Number of sources:} In the most detailed simulation to date, \citet{takami06err} finds a best fit to AGASA data on small scale anisotropy of UHECR arrival directions for a source density $n_s \equiv n_4 \times 10^{-4}\, {\rm Mpc}^{-3}$ with $n_4 \approx 1$.  This is consistent with \citet{augerLongAGN} determination that at least 61 sources contribute to the events above 57 EeV, 20 out of 27 of which correlate within $3.2^\circ$ with objects in the \citet{VCV} AGN catalog closer than 75 Mpc \citep{augerScience07} -- Auger08,07 below, respectively.   Assuming that all the uncorrelated events originate beyond 75 Mpc and that the $\approx$ 5 accidentally correlated events are representative in their distance distribution gives the weakest lower 
bound: $\geq$ 40 sources within 75 Mpc, corresponding $n_4 \gtrsim 0.3$.  Note that $n_s$ here is the number of sources visible in UHECRs which may be larger than the number visible via an accompanying photon jet, since magnetic deflection may un-beam UHECRs relative to photons.\\
{\it 3) Arrival time delay of UHECRs:}  The arrival time delay of a UHECR with net deflection $\delta \theta$ resulting from traversing a distance $D$ in a random weak magnetic field is $\tau \approx \frac{1}{2}\delta \theta^2 \, D/c$.  Assuming that the separations between the correlated Auger UHECRs and their candidate sources are due to random deflections in the extragalactic magnetic field, leads to an average time delay $ \langle \tau_{\rm delay} \rangle \equiv \tau_5 \times 10^{5}\, {\rm yr} \approx 10^5\, {\rm yr}$.

\section{Candidate UHECR accelerators}
\label{AGN,GRB}
The observational constraints above, combined with the results of Section \ref{params}, imply that conventional long-lived AGN jets (quasars, blazars, BLLacs,...) cannot be the primary site of UHECR acceleration.  There are too few AGN with the required luminosity within 75 Mpc in the Auger field of view and only a small fraction of the correlated VCV galaxies satisfy $L_{\rm bol} \geq 10^{45} \, E_{20}^2\, {\rm erg \, s^{-1}}$ \citep{zfg08}.  The possibility that the AGN which produced the observed UHECRs may no longer be so powerful  does not invalidate this argument:  if $N$ is the number of active sources within 75 Mpc required to account for the UHECRs observed today, then the number of sources expected at any randomly chosen time is also $N$.  Thus the probability of seeing none is miniscule since the expected number is $\gtrsim 40$.  Note that the time delay argument originally invoked by \citet{fp99} in their high-magnetic deflection scenario using Cen A {\it is} valid, since it was explicitly a single-source model; that scenario appears now to be excluded by the few-degree observed correlation between UHECR and many different sources, most far from Cen A (Auger07).

As pointed out by \citet{waxman95}, the conditions outlined in Section \ref{params} can be met by a GRB, if $\Gamma \approx 300$.  However the observational constraints make it difficult for GRBs to be the sole source of super-GZK cosmic rays. The local (isotropic equivalent) rate of long GRBs is \citep{gp07} $\Gamma_{\rm GRB} = (0.05 - 0.27) \, {\rm Gpc}^{-3}\,{\rm yr}^{-1} \equiv r_{10} 10^{-10 \pm 0.3}{\rm Mpc}^{-3}\,{\rm yr}^{-1} $. In order for GRB's to give the observed energy injection rate in UHECRs (\ref{Einj}), the energy per logarithmic interval in UHECRs produced by a typical GRB should be $10^{54.6}  \Gamma_{44.6}/r_{10}$ erg.  The spectrum of UHECRs produced by an individual source is expected to be and needs to be rather flat.  It is distributed over $ln(10^{9}) \approx 20$ logarithmic intervals, so the observed local GRB rate requires each GRB to contribute a total energy in cosmic rays of order $10^{56}  \Gamma_{44.6}/r_{10} $ erg.  The typical isotropic equivalent energy of photons from a GRB is $\approx {\rm few} \times10^{53}$ erg \citep {amati06}, so if GRBs are responsible for all the UHECRs, they must give a factor $\gtrsim 100$ times more energy to UHECRs than to photons.   Having a steeper UHECR spectrum is not a solution, because that would only exacerbate the difficulty of producing UHECRs at $10^{20}$ eV.  Nor is a flatter spectrum a solution because then there would be insufficient UHECRs at lower energies -- between the ankle and the GZK regime.   
Furthermore, there may be too few local GRBs.  Denoting the beaming factors for UHECRs and photons $b_{\rm UCR},\, b_\gamma$, the effective density of observable UHECR sources is $ (0.05-0.27)\, b_{\rm UCR} \, \tau_{\rm 5} \, 10^{-4}\, {\rm Mpc}^{-3}$.  Thus compatibility with observation is possible only if $b_{\rm UCR} \approx 1 \gg b_\gamma \approx 10^{-2}$.  
Additionally, for GRBs to be the sole source of UHECRs, the correlation with nearby AGN (Auger07) must be a statistical fluke or an induced correlation from the large scale structure of galaxies. 

\section{Giant AGN flares}
\label{AGNburst}

We will now argue that super-Eddington flares on super-massive black holes in (weak sub-Eddington) AGN are good candidates to be the source of UHECRs. The mass of an accretion disk near a black hole of mass $M$ (within twice the last stable orbit) is $\sim M_{\odot }\times (M/(10^7M_{\odot }))^{11/5}$, following \citet{ss73} and assuming an accretion rate equal to $10^{-3}$ of Eddington and alpha parameter $\alpha =10^{-2}$.  This would be a weak AGN, until a disk instability or tidal disruption of a passing star abruptly heats the disk.  With $\gsi  M_{\odot }$ in this hot disk within twice the last stable orbit, the accretion would be super-Eddington and one would conservatively expect flares of $\gsi 0.01 \, M_{\odot } c^2 = 2 \times 10^{52}$ erg.  

Without an accretion disk surrounding the BH, the time interval over which the stellar debris is accreted would be expected to be 10's to 100's of times the orbital return time, which is calculated to be $\sim 3 \times 10^6$s \citep{evansKochanek89,ulmer99}, leading to an event duration of order years and a luminosity too low in comparison to (\ref{lumi}) for the event to accelerate $10^{20}$ eV protons.  However if the BH is surrounded by an accretion disk (even a thin, cold one with little accretion taking place prior to the event), we expect the tidal debris and the accretion disk to disrupt one another in the first passage, leading to a thick disk and rapid accretion, within 10's of last-stable-orbit times.  Because the total amount of material involved in the accretion event is $\sim M_\odot$, residual accretion after the event is minimal, unlike in the case of galaxy mergers which power AGN that last $10^7$ yr or longer.

To give a concrete example, suppose that a burst producing a $10^{20}$ eV proton is due to a disk instability or to the tidal disruption of a Sun-like star, in an AGN with a $3\times 10^6\,M_{\odot }$ black hole.  The time scale of this mass injection (t, in the AGN frame) is uncertain, ranging from about $10^5$s for disk instabilities to $\approx 3\times 10^6$s for the rate of the tidal debris return \citep{evansKochanek89,ulmer99}.  
Suppose that $\Gamma  \approx 3$ and $B \approx 3$, then to satisfy both (\ref{conf}) and (\ref{phot}) we take $R \approx 3 \times 10^{16}$.  The corresponding variability time scale, $R\,  \Gamma^{-1} /c \approx 3 \times 10^5$ s, is comparable to or less than the duration of the event, as required\footnote{Choices of $\Gamma$ and $B$ implying a variability time that is longer than the duration of the event do not actually satisfy (\ref{conf}) and thus would not achieve the specified $E_{20}$. Decreasing $\Gamma$ below 3 can lead to problems with the variability time, for the mechanisms considered here, while increasing $\Gamma$ generally makes it easier to satisfy the requirements.  From (\ref{spec}), larger $\Gamma$ increases the luminosity and frequency range of photon emission.}.  With a power of $10^{45}$ erg/s and duration $t = 10^5$ to $3 \times 10^6$ s, the estimated energy is $10^{50}$ for a disk instability event or $3 \times 10^{51}$ erg for a tidal disruption event, comfortably compatible with the total energy budget in both cases.  

The stellar tidal disruption mechanism is particularly attractive because it is well-motivated theoretically and its predictions are relatively well-constrained.  The wait-time per galaxy for a stellar tidal disruption by its supermassive black hole, in the absence of an accretion disk,  is predicted to be $\approx 10^4 - 10^5$ yr \citep{magTremaine99}.  This estimate has received recent confirmation by \citet{gezari07} who have reported the UV and optical observation of two tidal disruption events in otherwise inactive galaxies; the detailed light curves and the inferred rate of tidal disruptions in inactive galaxies are consistent with theoretical predictions. The presence of an accretion disk may increase somewhat the rate per galaxy of stellar tidal disruptions, so we denote the wait-time between giant AGN flares as $T_4 10^4$ yr.  \citet{donley02}, comparing the ROSAT All Sky Survey and pointed observations, report evidence for soft x-ray flares of AGN whose estimated rate suggests $T_4 \approx 0.1 \, b_\gamma$, where $b_\gamma$ is the photon beaming fraction.  The relationship between the quiescent AGN luminosity and the conditions for a giant AGN flare is uncertain, so we take the density of galaxies than can host giant AGN flares to be $  f_2 \, 10^{-2} {\rm Mpc}^{-3}$, motivated by \citet{haoAGN05}'s determination that the density of AGN with $L[H_\alpha] > 10^5 L_{\odot }$ is $0.018\, {\rm Mpc}^{-3}$.  
Defining the rate of giant AGN flares to be $ \Gamma_{\rm GAF} =  r_6 \, 10^{-6} \,{\rm Mpc^{-3} yr^{-1}} $, observations thus give $r_6 = \, f_2/ T_4 \approx 20/b_\gamma$ and an apparent number density of UHECR sources $n_{4} = 0.1\,  r_6 \, \tau_5 \, b_{\rm UCR}/b_\gamma$, comfortably above the observational lower bound $n_{4} > 0.3$. 

Taking the cosmic ray spectrum to be flat over 20 logarithmic intervals and estimating the total energy in UHECRs in a stellar disruption-induced flare $\gtrsim 0.01 \, M_{\odot } c^2$,  gives $dE_{\rm UCR}/d ln E \equiv \epsilon_{51} 10^{51} $ erg, with $\epsilon_{51} \gtrsim 1$.  The predicted energy injection rate in UHECRs is thus also comfortably compatible with observation, reproducing  (\ref{Einj}) for $\epsilon_{51}\, r_6 = 0.7\, \Gamma_{44.6}$.  If disk instabilities are responsible for most UHECR-producing flares, the UHECR flux cannot be predicted but we can infer that the rate of instabilities needed to explain (\ref{Einj}) is about 30 times higher than in the tidal disruption case, since the total UHECR energy per event scales with the duration (to achieve the required minimum Poynting flux) .  The spectrum of milder disk-instabilities can indicate if this is reasonable.     

\section{Photon Bursts from AGN Flares}

The UHECR burst is accompanied by a photon burst of similar duration, observation of which will distinguish between a disk-instability and a tidal disruption origin.  Unlike for a GRB the accretion rate is not far above Eddington, so it is reasonable to expect emission from the accretion disk as in a quasar, with a total luminosity $\approx 10^{45}$ erg/s.  Thus an event capable of producing $10^{20}$ eV protons should be accompanied by a photon burst with a luminosity well in excess of the peak luminosity of supernovae ($\sim 10^{43}$ erg/s), with an isotropic component from the accretion disk and a beamed component produced in the jet that can only be seen if the jet happens to point toward us.  If equipartition obtains, the luminosity of the beamed component would be given by (\ref{lumi}) with spectrum (\ref{spec}), but since equipartition may not be reliable estimator, this is a less secure prediction than the bound on accretion luminosity.  

If AGN bursts are the sole source of UHECRs, we can estimate $p_{\rm GAF}$ -- the probability per unit volume that some galaxy is in a flaring state, i.e., the number density of Giant AGN Flares -- as follows.  Assuming the power in cosmic rays is an efficiency factor $f_{\rm CR}$ times the accretion power in photons, $L_{\rm GAF}$, we have   
\begin{eqnarray}
\label{PGAF}
p_{\rm GAF} & = & \Gamma_{\rm GAF}\,  t = \frac{\Gamma_{E ln E}\, \,\Delta ln E}{f_{\rm CR}\, L_{\rm GAF} }  \\ & = &3 \times 10^{-7} \, {\rm Mpc}^{-3} \left( \frac{ \Delta ln E}{20}\,\frac{ \Gamma_{44.6}}{1}  \frac{10^{45}}{f_{\rm CR} L_{\rm GAF} } \right). \nonumber
\end{eqnarray}
or about $\sim 3 \times 10^{-5} $ times the density of $L_*$ galaxies.  Note that if all UHECRs are produced by this AGN-burst mechanism, the probability of a galaxy being in the GAF state is independent of the flare mechanism (tidal disruption or disk instability) since the shorter duration and consequent lower energy injected per disk-instability event is exactly compensated by their greater required frequency.   SDSS has viewed of order $10^8$ galaxies at least twice, but has not compared the results of the different observations (M. Blanton, private communication).  A search of archival SDSS data is warranted.   Future large scale surveys like PANSTARS and LSST will see and measure the light curve of the predicted giant AGN flares, if they are as strong in the optical as suggested by (\ref{spec}).

AGN flares are in principle also observable by monitoring telescopes like Swift, Galex, and GLAST, although with a few exceptions only blazar-like objects have been detected at gamma-ray energies, so the effect of the beaming-factor $b_\gamma \gtrsim \Gamma^{-2}$ must be included when estimating the rate of detectable events.  GLAST observes the entire sky multiple times each day with the GBM (8 keV-30 MeV) and LAT (20 MeV - 300 GeV).  Denoting the beamed SED $\nu L_{\nu, \rm GAF}$ and the 1-day sensitivity $(\nu F_\nu)_{\rm sen}$, and using (\ref{PGAF}), the number of observable flares per year is
\begin{equation}
\label{detectableGAFrate}
N_{\rm GAF,yr} \approx  30 \, \frac{b_\gamma }{0.1} \left( \frac{  \nu L_{\nu, \rm GAF} }{10^{45} } \frac{10^{-9}  }{ (\nu F_\nu)_{\rm sen} } \right)^\frac{3}{2}  t_d^{-\frac{1}{4}} \frac{p_{\rm GAF}}{ 3 \times 10^{-7}},
\end{equation}
where the duration of the event is $t_d $ days, with $t_d \approx 1-30$ for tidal disruption and stellar tidal disruption bursts respectively.
  
Thus if AGN bursts are responsible for UHECRs, we predict that GLAST will observe large numbers of these bursts each year.  The flare duration should be long enough in either the tidal disruption or disk instability case, that GLAST can issue an alert to allow detailed observation at other wavelengths.  The spectra of individual flares and the distribution of maximum energies among different flares will elucidate the acceleration environment by comparing reality with the equipartition estimate (\ref{spec}).

\section{UHECR predictions}
Unlike conventional AGN jets or GRBs, giant AGN flares have no difficulty explaining the UHECR observations discussed in Section \ref{obs}.  The cosmic ray energy injection rate is a prediction of the tidal-disruption scenario and explains the observed rate (\ref{Einj}) for reasonable parameter values.  Furthermore,  the effective density of visible UHECR sources is estimated to be well above the observational lower limit and can be larger if UHECRs are not so beamed as the photons from the jet.  

The giant AGN flare model makes further predictions that are supported by UHECR observations.  The correlation between UHECRs and AGN observed by Auger is of course a basic prediction of this scenario, but a further prediction is that the correlating AGN -- observed today with photons -- are not especially luminous, since after a flare the AGN quickly returns to a weakly accreting state as discussed in Section \ref{AGNburst}.  Furthermore, when the UHECR arrival time delay is short compared to the time since the last flare,  the spectrum should be characteristic of a bursting source rather than of a continuous source, modulo GZK distortions \citep{waxmanME,gfICRC07}.  In fact, the spectrum of the Ursa Major cosmic ray cluster -- 4-5 events in the combined AGASA-HiRes data \citep{HRGF,gfclus} which is the only example so far of a cluster of events from a single source -- favors a bursting over a continuous source \citep{gfICRC07}.  These and other consequences of UHECRs being produced by giant AGN flares will be discussed in greater detail elsewhere.  

\section{Summary}
Theoretical and observational constraints on sources of ultrahigh energy protons are derived.  Conventional, powerful AGN fail on the grounds of being too rare, and GRBs are only viable as the sole source of UHECRs if their energy in cosmic rays is more than two orders of magnitude greater than in photons.  Powerful, short duration bursts of AGN, predicted to be induced by the tidal disruption of a star or an instability in the accretion disk, are shown to satisfy all of the requirements for UHECR acceleration.  Numerical simulations of stellar tidal disruption in the presence of a thin accretion disk are needed, to test our prediction of efficient and rapid generation of a thick disk.  The photon counterparts should be observable by GLAST and in archival SDSS data or the next generation of optical surveys.   

\acknowledgements
We thank A. Berlind, P. Biermann, M. Blanton, A. Fillipenko, P. Goldreich, D. Hogg, S. Kulkarni, A. MacFadyen, J. McEnery, T. Piran, M. Strauss,  E. Waxman and especially J. Gelfand, for information and discussions. GRF thanks her collaborators in the Pierre Auger Observatory for stimulating and enlightening discussions and comments.  This research has been supported by NSF-PHY-0401232 and PHY-0701451, and by the David and Lucile Packard Foundation.  


\begin{thebibliography}{22}
\expandafter\ifx\csname natexlab\endcsname\relax\def\natexlab#1{#1}\fi

\bibitem[{Abbasi {et~al.}(2005)}]{HRGF}
Abbasi, R.~U. {et~al.} 2005, Astrophys. J., 623, 164

\bibitem[{{Amati}(2006)}]{amati06}
{Amati}, L. 2006, MNRAS, 372, 233

\bibitem[{{Berezinsky}(2008)}]{berez07}
{Berezinsky}, V. 2008, Advances in Space Research, 41, 2071

\bibitem[{{Blandford} \& {Znajek}(1977)}]{bz77}
{Blandford}, R.~D. \& {Znajek}, R.~L. 1977, \mnras, 179, 433

\bibitem[{{Donley} {et~al.}(2002){Donley}, {Brandt}, {Eracleous}, \&
  {Boller}}]{donley02}
{Donley}, J.~L., {Brandt}, W.~N., {Eracleous}, M., \& {Boller}, T. 2002, \aj,
  124, 1308

\bibitem[{Evans \& Kochanek(1989)}]{evansKochanek89}
Evans, C.~R. \& Kochanek, C.~S. 1989, Astrophys. J., 346, L13

\bibitem[{Farrar(2005)}]{gfclus}
Farrar, G.~R. 2005, astro-ph/0501388

\bibitem[{Farrar(2007)}]{gfICRC07}
---. 2007, ICRC 2007 and arXiv:0708.1617 [astro-ph]

\bibitem[{Farrar \& Piran(2000)}]{fp99}
Farrar, G.~R. \& Piran, T. 2000, Phys. Rev. Lett., 84, 3527

\bibitem[{Gezari {et~al.}(2007)}]{gezari07}
Gezari, S. {et~al.} 2007, Astrophys. J. in press

\bibitem[{{Guetta} \& {Piran}(2007)}]{gp07}
{Guetta}, D. \& {Piran}, T. 2007, Journal of Cosmology and Astro-Particle
  Physics, 7, 3

\bibitem[{Hao {et~al.}(2005)}]{haoAGN05}
Hao, L. {et~al.} 2005, "Astronom. J.", 129, 1795

\bibitem[{Magorrian \& Tremaine(1999)}]{magTremaine99}
Magorrian, J. \& Tremaine, S. 1999, MNRAS, 309, 447

\bibitem[{{Shakura} \& {Sunyaev}(1973)}]{ss73}
{Shakura}, N.~I. \& {Sunyaev}, R.~A. 1973, Astron. Astrophys., 24, 337

\bibitem[{{Takami} {et~al.}(2006){Takami}, {Yoshiguchi}, \&
  {Sato}}]{takami06err}
{Takami}, H., {Yoshiguchi}, H., \& {Sato}, K. 2006, \apj, 653, 1584

\bibitem[{{The Pierre Auger Collaboration}(2007)}]{augerScience07}
{The Pierre Auger Collaboration}. 2007, Science, 318, 939

\bibitem[{{The Pierre Auger Collaboration}(2008)}]{augerLongAGN}
---. 2008, Astroparticle Physics, 29, 188

\bibitem[{Ulmer(1999)}]{ulmer99}
Ulmer, A. 1999, Astrophys. J., 514, 180

\bibitem[{{V{\'e}ron-Cetty} \& {V{\'e}ron}(2006)}]{VCV}
{V{\'e}ron-Cetty}, M.-P. \& {V{\'e}ron}, P. 2006, "Astronom. \& Astrophys. ",
  455, 773

\bibitem[{Waxman(1995)}]{waxman95}
Waxman, E. 1995, Phys. Rev. Lett., 75, 386

\bibitem[{Waxman \& Miralda-Escude(1996)}]{waxmanME}
Waxman, E. \& Miralda-Escude, J. 1996, Astrophys. J., 472, L89

\bibitem[{Zaw {et~al.}(2008)Zaw, Farrar, \& Greene}]{zfg08}
Zaw, I., Farrar, G.~R., \& Greene, J. , arXiv:0806.3470.

\end{thebibliography}

\end{document}